\documentclass[11pt]{amsart}

\usepackage[body={16cm,23cm}]{geometry}
\usepackage{amsmath,amssymb}
\usepackage{amsfonts}
\usepackage[mathscr]{eucal}
\usepackage{epic}
\usepackage{eepic}

\newcommand{\ds}{\displaystyle}
\def\EXP{\textrm{{\large e}}}
\def\eqref#1{(\ref{#1})}
\newcommand{\uop}{\mathbf{u}}
\newcommand{\vop}{\mathbf{v}}

\newcommand{\uo}{{\boldsymbol{\lambda}}}
\newcommand{\vo}{{\boldsymbol{\mu}}}
\newcommand{\xop}{\mathbf{x}}
\newcommand{\yop}{\mathbf{y}}

\newcommand{\Nop}{\mathbf{h}}

\def\tvec#1#2{\left(\begin{array}{ll} #1 \\ #2 \end{array}\right)}
\begin{document}

\title{Ansatz of Hans Bethe for a two-dimensional Bose gas}

\author{S. Sergeev}
\email{Sergey.Sergeev@anu.edu.au}
\address{Department of Theoretical Physics,\\
         Research School of Physical Sciences and Engineering,
    Australian National University, Canberra, ACT 0200, Australia.}

\thanks{The work was supported by the Australian Research Council}

\begin{abstract}
The method of $q$-oscillator lattices, proposed recently in
[hep-th/0509181], provides the tool for a construction of various
integrable models of quantum mechanics in $2+1$ dimensional
space-time. In contrast to any one dimensional quantum chain, its
two dimensional generalizations -- quantum lattices -- admit
different geometrical structures. In this paper we consider the
$q$-oscillator model on a special lattice. The model may be
interpreted as a two-dimensional Bose gas. The most remarkable
feature of the model is that it allows the coordinate Bethe
Ansatz: the $p$-particles' wave function is the sum of plane
waves. Consistency conditions is the set of $2p$ equations for $p$
one-particle wave vectors. These ``Bethe Ansatz'' equations are
the main result of this paper.
\end{abstract}

\maketitle

\section{Introduction}

Integrable models of quantum mechanics in discrete space-time
describe a system of interacting quantum observables (spin
operators, oscillators, etc.) situated in sites of an one
dimensional chain (the models in $1+1$ dimensional space-time), or
in the vertices of a two dimensional lattice (models in $2+1$
dimensional space-time). The subject of this paper is the second
variant -- the quantum integrable lattices.

Formulation of a \emph{realistic} integrable model of quantum
mechanics in $2+1$ dimensional space-time was a long-standing
problem in the theory of integrable systems. Examples of higher
dimensional integrable systems are known, but the integrability
seemed to be a very high price paid to the detriment of a physical
interpretation.

Both known classes of integrable quantum lattices may be specified
by the definition of the algebra of observables. In contrast to
the quantum chains, where the algebra of observables may be e.g.
an evaluation representation of any quantum group, in the case of
quantum lattices the algebra of observables may be either the
local\footnote{The term ``local'' means that the quantum
observables in different vertices commute, as well as it is in the
quantum chains case} Weyl algebra \cite{Sergeev:2000tmp} or the
local $q$-oscillator algebra \cite{Sergeev:2005b}. Other examples
of integrable quantum lattices are not known.

When the algebra of observables (and its representation) is fixed,
there exists an almost unique way to produce the key notion of the
integrability -- a complete set of commutative operators. For the
spin chains it is the way of Lax operators and transfer matrices
with e.g. fundamental representations in the auxiliary space. In
both classes of quantum lattices the integrals of motion may be
produced by a decomposition of a determinant of a certain big
operator-valued matrix with respect to two spectral parameters
\cite{Sergeev:1999jpa,Sergeev:2005d}. Therefore the integrals of
motion rather have a combinatorial nature. Both classes of quantum
lattices allow a definition of another kind of operators: the
layer-to-layer transfer matrices, related to intertwining
relations, tetrahedron equations etc. Such
representation-dependent layer-to-layer transfer matrices are the
kind of discrete-time evolution operators for the quantum
lattices, their structure is much more complicated, but since
their integrals of motion are known, it is not necessary to
consider them at the first step.

The concept of quantum lattices implies a certain feature which is
not applicable to the quantum chains. One may vary not only a size
of the lattice, but its geometry and topology as well. This is the
reason why we discuss the \emph{classes} of models. The methods of
Weyl algebra and $q$-oscillator algebra are the frameworks for
construction of various quantum lattices.

Mainly the simplest geometry was considered before -- the square
lattices with periodical boundary conditions\footnote{Examples of
Weyl algebra non-square lattices were mentioned in
\cite{Sergeev:2004pn}. In particular, the Weyl algebra framework
on the different shapes of the lattices produces the relativistic
Toda chain as well as the quantum discrete Liouville model.}. The
simple square lattice with the sizes $n\times m$ may be regarded
as the length-$m$ chain of its length-$n$ lines, and therefore
such quantum lattice is effectively a quantum chain. In
particular, the quantum lattice with the Weyl algebra at a root of
unity corresponds to auxiliary transfer matrices
\cite{Sergeev:2000tmp} of the $\mathcal{U}_q(\widehat{gl}_n)$
generalized chiral Potts model \cite{Bazhanov:1990qk,Date:1990bs}.
The $q$-oscillator quantum lattice corresponds to the auxiliary
transfer matrices for reducible oscillator representation of
$\mathcal{U}_q(\widehat{gl}_n)$ \cite{Sergeev:2005b}. From the
point of view of quantum mechanics in $2+1$ dimensional
space-time, the square lattices provide a non-realistic models
since the translation operators do not belong to the set of
integrals of motion and therefore momenta of eigenstates can not
be defined. In the quantum chain interpretation, it means that
$U(1)$ charges are the variables, and the chain is homogeneous
only on subspaces where all $U(1)$ charges are the same in all
sites. Otherwise, the translation invariance is lost.

The issue is evident. One has to consider a lattice which is not a
chain of its lines. We define such lattice in the next sections.
The framework of $q$-oscillator model on our special lattice
immediately produces a kind of two dimensional Bose gas with
physical dynamics. The Fock vacuum is the natural reference state,
the total $q$-oscillators occupation number -- the number of
bosons -- is conserved, and the $p$-particles wave function is the
superposition of one-particle plane waves. Each plane wave is
characterized by two components of its momentum, and the
consistency conditions give $2p$ equations for $p$ two-components
momenta. These consistency conditions look much more complicated
than the Bethe Ansatz equations for a quantum chain, we do not
investigate them in details here. The aim of this paper is just to
present the method, the model, and the ``Bethe Ansatz'' equations.

\section{Framework of $q$-oscillator algebra.}

We start with the formulation of a generic $q$-oscillator lattice.

Let $\mathcal{L}$ be a lattice formed by a number of directed
lines on a torus. Pairwise intersections of the lines are the
vertices of the lattice. Let the vertices are enumerated in some
way, for instance by the integer index $j=1,2,3,\dots,\Delta$,
where $\Delta$ is the number of vertices.

\begin{figure}[ht]
\setlength{\unitlength}{0.25mm}
\begin{picture}(400,100)
\put(0,50){\begin{picture}(50,50) \linethickness{0.1mm}
\put(0,25){\vector(1,0){50}} \put(25,0){\vector(0,1){50}}
\put(22,-30){\small $1$}
\end{picture}}
\put(70,50){\begin{picture}(50,50) \linethickness{0.5mm}
\put(0,26){\line(1,0){24}}\put(24,25){\vector(0,1){24}}
\put(26,0){\line(0,1){24}}\put(26,24){\vector(1,0){24}}
\put(22,-30){\small $\nu_j^2$}
\end{picture}}
\put(140,50){\begin{picture}(50,50)
\linethickness{0.1mm}\put(0,25){\vector(1,0){50}}
\linethickness{0.5mm}\put(25,0){\vector(0,1){50}}
\put(22,-30){\small $\lambda_j q^{\Nop_j}$}
\end{picture}}
\put(210,50){\begin{picture}(50,50)
\linethickness{0.5mm}\put(0,25){\vector(1,0){50}}
\linethickness{0.1mm}\put(25,0){\vector(0,1){50}}
\put(22,-30){\small $\mu_j q^{\Nop_j}$}
\end{picture}}
\put(280,50){\begin{picture}(50,50)
\linethickness{0.1mm}\put(0,25){\line(1,0){25}}\put(25,25){\vector(0,1){25}}
\linethickness{0.5mm}\put(25,0){\line(0,1){25}}\put(25,25){\vector(1,0){25}}
\put(22,-30){\small $\nu_j\xop_j$}
\end{picture}}
\put(350,50){\begin{picture}(50,50)
\linethickness{0.5mm}\put(0,25){\line(1,0){25}}\put(25,25){\vector(0,1){25}}
\linethickness{0.1mm}\put(25,0){\line(0,1){25}}\put(25,25){\vector(1,0){25}}
\put(20,-30){\small $\nu_j\yop_j$}
\end{picture}}
\end{picture}
\caption{The ``weights'' $f_j$ of $j^{\;\textrm{th}}$ vertex,
$j=1,...,\Delta$.} \label{fig-weights}
\end{figure}
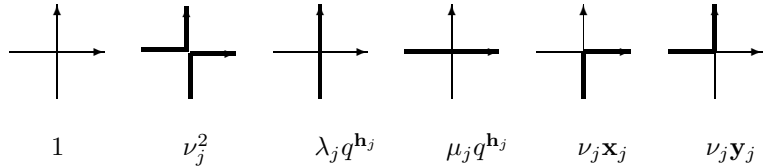

The $q$-oscillator generators $\xop_j,\yop_j$ and $\Nop_j$ are
assigned to vertex $j$. The oscillators for the different vertices
commute. Locally, we define the $q$-oscillator algebra by
\begin{equation}\label{alg}
\xop_j\yop_j\;=\; 1-q^{2+2\Nop_j}\;,\quad \yop_j\xop_j = 1-
q^{2\Nop_j}\;,\quad \xop_jq^{\Nop_j}=q^{\Nop_j+1}\xop_j\;,\quad
\yop_j q^{\Nop_j} = q^{\Nop_j-1}\yop_j\;.
\end{equation}
Consider next the system of non-self-intersecting paths along the
edges of the lattice. The lines of the lattice are directed, and
we demand that paths must follow the orientation of the edges. The
paths may go through a single vertex in one of six variants shown
in Fig.\ref{fig-weights}.
The vertex ``weight'' $f_j$ is associated with each variant of
bypassing as it is shown in Fig. \ref{fig-weights}. Coefficients
$\lambda_j$, $\mu_j$ and $\nu_j$ are extra vertex
$\mathbb{C}$-valued parameters, they are related by $\nu_j^2\equiv
-q^{-1}\lambda_j\mu_j$. In general, they are free up to a single
condition: if two vertices $j$ and $j'$ are formed by the
intersections of the same lines, their $\mathbb{C}$-valued
parameter must be the same \cite{Sergeev:2005d}.

For any path $P$ let
\begin{equation}\label{tP}
\mathbf{t}_{P}\;=\;\prod_{\textrm{along } P} \; f_j\;.
\end{equation}
Note, if a path does not touch a vertex, then it vertex's
contribution to $\mathbf{t}_P$ is just the unity according to the
leftmost variant of Fig. \ref{fig-weights}. Recall in addition,
the $q$-oscillator algebra is local, elements of different
vertices commute. Therefore the notion of the product along the
path in (\ref{tP}) is well defined.

Let $A$ and $B$ be the two basic homotopy cycles of the torus. Any
path has a certain homotopy class, $P\sim nA+mB$. Let
\begin{equation}\label{IM}
\mathbf{t}_{n,m}\;=\;\sum_{P\;:\; P\sim nA+mB} \mathbf{t}_P\;.
\end{equation}
Here the sum is taken over all possible paths of homotopy class
$nA+mB$. Formula (\ref{IM}) ends the formulation of $q$-oscillator
lattice's prescription. The point is that for \emph{any} lattice
$\mathcal{L}$, formed by directed lines on the torus, the
operators $\mathbf{t}_{n,m}$ form the commutative set:
\begin{equation}\label{CM}
\mathbf{t}_{n,m}\;\mathbf{t}_{n',m'}\;=\;
\mathbf{t}_{n',m'}\;\mathbf{t}_{n,m}\qquad \forall \quad
n,m,n',m'\;.
\end{equation}
Moreover, if there are no lines of trivial homotopy class in
$\mathcal{L}$, the set of integrals of motion is complete.

Integrals of motion may be combined into a polynomial of two
variables,
\begin{equation}\label{TM}
\mathbf{T}(u,v)\;=\;\sum_{n,m} u^n v^m \mathbf{t}_{n,m}\;.
\end{equation}
Operator $\mathbf{T}$ has the structure of the layer-to-layer
transfer matrix, equations (\ref{CM}) are equivalent to the
commutativity of transfer matrices with different spectral
parameters $u,v$. The commutativity may be proven for simple
lattices with the help of tetrahedron equation
\cite{Bazhanov:1981zm,Sergeev:2005b,Sergeev:2005c}.

\section{The lattice}

\begin{figure}[ht]
\setlength{\unitlength}{0.25mm}
\begin{picture}(500,350)
\put(50,25){
\begin{picture}(400,300) 
\thinlines \dashline[-10]{5}(0,0)(400,0)(400,300)(0,300)(0,0)
\path(40,0)(0,120) \path(140,0)(40,300) \path(240,0)(140,300)
\path(340,0)(240,300) \path(400,120)(340,300)
\path(0,75)(300,0) \path(0,175)(400,75) \path(0,275)(400,175)
\path(300,300)(400,275)
%
\put(55,267){\scriptsize $1$} \put(92,157){\scriptsize $2$}
\put(129,50){\scriptsize $3$}
\put(240,20){\scriptsize $M+3$} \put(202,130){\scriptsize $M+2$}
\put(164,240){\scriptsize $M+1$}

\end{picture}}
\put(0,162){$A$}\put(5,185){$\uparrow$}
\put(250,0){$B\;\rightarrow$}
\end{picture}
\caption{The lattice formed by the lines $\alpha\sim 4A-B$ and
$\beta\sim 3B-A$.}\label{fig-1}
\end{figure}
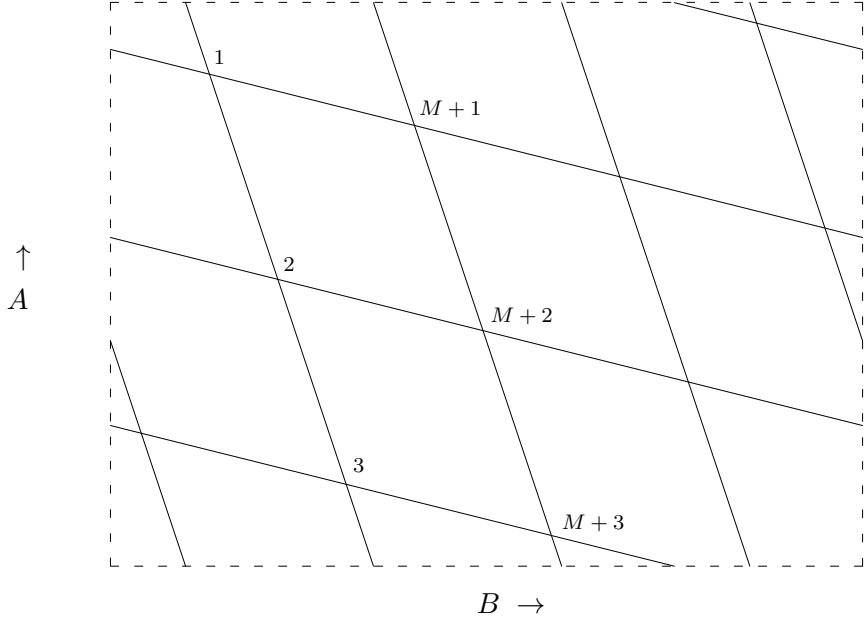

Turn now to the definition of our special lattice. It is formed by
the intersections of only two lines $\alpha$ and $\beta$. With
respect to the basic cycles $A$ and $B$ of the torus, the lines
have the classes
\begin{equation}\label{idea}
\alpha\;\sim\;N A - B\;,\qquad \beta\;\sim\; M B - A\;.
\end{equation}
Example of such lattice is given in Fig. \ref{fig-1}. There the
opposite dashed borders are identified (dashed lines are the cuts
of the torus). The line $\alpha$ is directed up (along the
$A$-cycle), the line $\beta$ is directed to the right (along the
$B$-cycle).
The lines are intersecting in
\begin{equation}
\Delta\;=\; NM - 1
\end{equation}
points.

We enumerate the vertices by number $j$, $j\in\mathbb{Z}_\Delta$.
The numeration is successive along the reverse direction of
$\alpha$-line, see Fig. \ref{fig-1}. The shift $j\to j-1$
corresponds to a one-step translation along the $\alpha$-line,
whereas the sift $j\to j+M$ corresponds to the one-step
translation along the $\beta$-line.

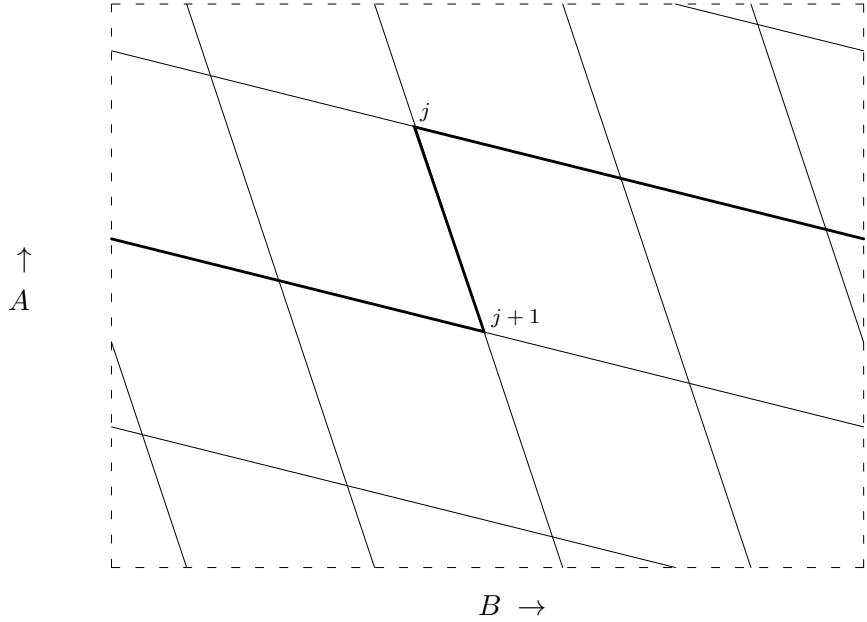
\begin{figure}[ht]
\setlength{\unitlength}{0.25mm}
\begin{picture}(500,350)
\put(50,25){
\begin{picture}(400,300) 
\thinlines \dashline[-10]{5}(0,0)(400,0)(400,300)(0,300)(0,0)
\path(40,0)(0,120) \path(140,0)(40,300) \path(240,0)(140,300)
\path(340,0)(240,300) \path(400,120)(340,300)
\path(0,75)(300,0) \path(0,175)(400,75) \path(0,275)(400,175)
\path(300,300)(400,275)
%
\put(202,130){\scriptsize $j+1$} \put(164,240){\scriptsize $j$}
\Thicklines
\path(0,175)(198.18,125.45)(161.18,234.55)(400,175)
\end{picture}}
\put(0,162){$A$}\put(5,185){$\uparrow$}
\put(250,0){$B\;\rightarrow$}
\end{picture}
\caption{Path of the class $B$ -- an element of
$\mathbf{t}_{0,1}$.}\label{fig-2}
\end{figure}

Now the lattice is formulated, and combinatorial rules of Fig.
\ref{fig-weights} may be applied. Note, since all the vertices of
the lattice are formed by the same lines, all their
$\mathbb{C}$-valued parameters are the same, and one may put
\begin{equation}\label{lambda-mu}
\lambda\;=\;\mu\;=\;1\;,\quad \nu^2=-q^{-1}\;.
\end{equation}
Generating function (\ref{TM}) has the structure
\begin{equation}\label{our-TM}
\mathbf{T}(u,v)=u^Nv^{-1} q^{\mathcal{N}} + v^Mu^{-1}
q^{\mathcal{N}} + \sum_{n=0}^{N-1} \sum_{m=0}^{M-1} u^n v^m
\mathbf{t}_{n,m}\;,
\end{equation}
where $\mathcal{N}$ is the total occupation number
\begin{equation}
\mathcal{N}= \sum_j\Nop_j\;.
\end{equation}
Elements $q^{\mathcal{N}}$ in (\ref{our-TM}) are the values of
both $\mathbf{t}_{N,-1}$ and $\mathbf{t}_{-1,M}$, combinatorially
they are the paths along all $\alpha$ or $\beta$ lines (the third
and the fourth variants of Fig. \ref{fig-weights} for all
vertices). Elements $\mathbf{t}_{0,0}$ and $\mathbf{t}_{N-1,M-1}$
correspond to the empty path and the complete path (the first and
the second variants of Fig. \ref{fig-weights} for all vertices)
correspondingly, their values are
\begin{equation}
\mathbf{t}_{0,0}=1\;,\quad
\mathbf{t}_{N-1,M-1}=(-q^{-1})^{\Delta}\;.
\end{equation}
In what follows, we will consider the elements $\mathbf{t}_{0,1}$
and $\mathbf{t}_{1,0}$ given by
\begin{equation}\label{t01}
\begin{array}{l}
\ds \mathbf{t}_{1,0}\;=\;-q^{-1}\sum_j \xop_j\yop_{j+M}
q^{\Nop_{j+1}+\Nop_{j+2}+\cdots + \Nop_{j+M-1}}\;,\\
\\
\ds \mathbf{t}_{0,1}\;=\;-q^{-1}\sum_j \xop_j\yop_{j+1}
q^{\Nop_{j+M}+\Nop_{j+2M}+\cdots + \Nop_{j+(N-1)M}}\;.
\end{array}
\end{equation}
Combinatorial summand of $\mathbf{t}_{0,1}$ is shown in Fig.
\ref{fig-2}.

\section{Eigenstates.}

The total occupation number is the conserving quantity, and
therefore we may construct the eigenstates of the model
step-by-step just increasing the number of bosons for the Fock
space representation of the $q$-oscillator algebra. The Fock
vacuum $|0\rangle$ is defined by
\begin{equation}\label{vacuum}
\xop_j\;|0\rangle\;=\;0\;,\quad \Nop_j\;|0\rangle \;=\;0\quad
\forall\ j\in\mathbb{Z}_\Delta\;.
\end{equation}
The Fock vacuum is evidently the eigenstate of the model,
eigenvalues of all $\mathbf{t}_{n,m}$ are zeros (except
$\mathbf{t}_{0,0}$, $\mathbf{t}_{N-1,M-1}$ and $q^{\mathcal{N}}$).

In what follows, we concentrate on the diagonalization of
$\mathbf{t}_{0,1}$. Its eigenstates are defined uniquely. They
will be evidently the eigenstates of $\mathbf{t}_{1,0}$ and, due
to the uniqueness, they must be the eigenstates of all the other
integrals of motion. We will use below the normalized form of
(\ref{t01}):
\begin{equation}\label{norm}
\boldsymbol{\tau}_\alpha \;=\; -\frac{q}{1-q^2} \;
\mathbf{t}_{0,1}\;,\quad \boldsymbol{\tau}_\beta \;=\;
-\frac{q}{1-q^2}\; \mathbf{t}_{1,0}\;.
\end{equation}

\subsection{One boson state}

Consider the states with the single boson, $\mathcal{N}=1$. Let
\begin{equation}
|\phi_j\rangle \;=\; \yop_j \; |0\rangle
\end{equation}
be the basis of one-boson states. On this subspace both operators
(\ref{norm}) are the translation operators:
\begin{equation}
\boldsymbol{\tau}_\alpha\;|\phi_j\rangle \;=\; \;
|\phi_{j+1}\rangle\;,\quad \boldsymbol{\tau}_\beta\;|\phi_j\rangle
\;=\; \; |\phi_{j+M}\rangle\;.
\end{equation}
Therefore
\begin{equation}\label{psi-one}
|\Psi\rangle\;=\;\sum_{j} \; \omega^j \; |\phi_j\rangle\;,\quad
\omega^\Delta=1\;,
\end{equation}
is the eigenstate of the model, the eigenvalues of (\ref{norm})
are
\begin{equation}\label{tau-one}
\boldsymbol{\tau}_\alpha \; |\Psi\rangle \;=\; |\Psi\rangle \;
\omega^{-1}\;,\quad \boldsymbol{\tau}_\beta\; |\Psi\rangle \;=\;
|\Psi\rangle \; \omega^{-M}\;.
\end{equation}
Since $\omega$: $\omega^\Delta=1$ may take $\Delta$ different
values, the set of (\ref{psi-one}) is complete on the subspace
$\mathcal{N}=1$.

Formally, the spectrum of both translation operators
(\ref{tau-one}) is defined by one parameter $\omega$. Note,
eigenvalues $\omega^{-1}$ and $\omega^{-M}$ are dual in the sense
$(\omega^{-M})^{N}=\omega^{-1}$. When $N,M\to \infty$, one may
talk about two independent components of momentum. Namely, let
\begin{equation}\label{k-1}
k\;=\; N\; k_\alpha + k_\beta\;,
\end{equation}
where $k_\alpha$ and $k_\beta$ are relatively small. Then, when
$N,M\to \infty$,
\begin{equation}\label{tr-1}
\omega^{-1}\;=\;\EXP^{2\pi i k/\Delta}\;\to\; \EXP^{2\pi i
k_\alpha/M}\;,\quad \omega^{-M} \;=\; \EXP^{2\pi i
Mk/\Delta}\;\to\; \EXP^{2\pi i k_\beta/N}\;,
\end{equation}
i.e. in the thermodynamical limit $2\pi k_\alpha/M$ and $2\pi
k_\beta/N$ play the r\^oles of independent components of the
momentum.

\subsection{Two bosons state}

Turn next to $\mathcal{N}=2$. Define
\begin{equation}
c_{j,k}\;=\; \delta_{j,k+M}+\delta_{j,k+2M}+\cdots +
\delta_{j,k+(N-1)M}\;.
\end{equation}
Then
\begin{equation}
\begin{array}{l}
\ds \boldsymbol{\tau}_\alpha\;\cdot\; \yop_j \yop_k |0\rangle
\;=\; (q^{c_{k,j}}\yop_{j+1}\yop_k+q^{c_{j,k}}\yop_j\yop_{k+1})
|0\rangle\;,\quad j\neq k\;,\\
\ds \boldsymbol{\tau}_\alpha \;\cdot\; \yop_k^2 |0\rangle \;=\;
(1+q^2)\yop_{k+1}\yop_k|0\rangle\;.
\end{array}
\end{equation}
Consider a two-particles state with zero momentum of the mass
center:
\begin{equation}
|\Psi\rangle = \sum_{j}\left( \Psi_0 (1+q^2)^{-1}\yop_j^2|0\rangle
+ \sum_{k > 0} \Psi_k \yop_j\yop_{j+k}|0\rangle\right)
\end{equation}
Eigenvalue equation $\boldsymbol{\tau}_\alpha \;|\Psi\rangle =
|\Psi\rangle \; \tau_\alpha$ reads in components
\begin{equation}\label{seq}
\begin{array}{l}
 \ds \tau_\alpha\Psi_0=(1+q^2)\Psi_1\;, \\
 \ds \tau_\alpha\Psi_{nM-1}=q\Psi_{nM} +\Psi_{nM-2}\;,\\
 \ds \tau_\alpha\Psi_{nM}=\Psi_{nM+1} + \Psi_{nM-1}\;,\\
 \ds \textrm{in all other cases}:\;\;\;
 \tau_\alpha\Psi_k=\Psi_{k-1}+\Psi_{k+1}\;.
\end{array}
\end{equation}
Periodicity condition is simply
\begin{equation}
\Psi_{k}=\Psi_{\Delta-k}\;.
\end{equation}
Define the double index: $j=Mn+m$ $\to$ $j=(n,m)$,
\begin{equation}\label{double}
\Psi_{Mn+m}\;=\;\Psi_{(n,m)}\;,\quad 0\leq m < M\;, \;\;\; 0\leq n
< N\;.
\end{equation}
Equations (\ref{seq}) have in the general position the form
$\tau_\alpha\Psi_{(n,m)}=\Psi_{(n,m-1)}+\Psi_{(n,m+1)}$.
Therefore,
\begin{equation}\label{mansatz}
\Psi_{(n,m)} \;=\; P_n\omega^m + Q_n \omega^{-m}\;,\quad
\tau_\alpha = \omega + \omega^{-1}\;.
\end{equation}
Initial conditions (the first of (\ref{seq})) give
\begin{equation}\label{PQ}
\frac{Q_0}{P_0}=\frac{1-q^2\omega^2}{q^2-\omega^2}\;.
\end{equation}
The second and third relations of (\ref{seq}) give
\begin{equation}\label{PQ-matrix}
\tvec{P_n}{Q_n} \;=\;
\mathcal{M}\;\cdot\;\tvec{P_{n-1}}{Q_{n-1}}\;,\quad
\mathcal{M}\;=\;\left(\begin{array}{cc} \ds \omega^M
\frac{q^2-\omega^2}{q(1-\omega^2)} & \ds
\omega^{-M}\frac{\omega^2(q^2-1)}{q(1-\omega^2)}\\
\ds \omega^M \frac{1-q^2}{q(1-\omega^2)} & \ds \omega^{-M}
\frac{1-q^2\omega^2}{q(1-\omega^2)}\end{array} \right) \;.
\end{equation}
Let $\Omega$ and $\Omega^{-1}$ be the eigenvalues of
$\mathcal{M}$. Then one may choose the basis of eigenvectors of
$\mathcal{M}$ such that
\begin{equation}
P_n\;=\;A_{++}\Omega^n\;+\; A_{-+}\Omega^{-n}\;,\quad Q_n\;=\;
A_{+-}\Omega^n \;+\; A_{--}\Omega^{-n}\;.
\end{equation}
The boundary condition $\Psi_k=\Psi_{\Delta-k}$ reads in
double-indices $\Psi_{(N-1,m)}=\Psi_{(0,M-1-m)}$, i.e.
\begin{equation}\label{MPQ}
\tvec{P_{N-1}}{Q_{N-1}} \;=\;
\mathcal{M}^{N-1}\;\cdot\;\tvec{P_0}{Q_0} \;=\;
\tvec{\omega^{1-M}Q_0}{\omega^{M-1}P_0}\;.
\end{equation}
The following par of equations summarizes all the calculations:
\begin{equation}\label{BAE2}
\begin{array}{l}
\ds \Omega + \Omega^{-1} \;=\; \omega^M
\frac{q^2-\omega^2}{q(1-\omega^2)} \;+\; \omega^{-M}
\frac{1-q^2\omega^2}{q(1-\omega^2)}\;,\\
\\
\ds \omega+\omega^{-1} = \Omega^N
\frac{q^2-\Omega^2}{q(1-\Omega^2)} \;+\; \Omega^{-N}
\frac{1-q^2\Omega^2}{q(1-\Omega^2)}\;.
\end{array}
\end{equation}
Here the first equation is the definition of $\Omega$ (it is the
characteristic polynomial of $\mathcal{M}$). The second equation
comes from (\ref{MPQ}) after excluding of $P_0$, $Q_0$ via
(\ref{PQ}). The final answer is the Bethe Ansatz (two-particles
wave function is the superposition of one-particle plane waves)
\begin{equation}\label{2wave}
\Psi_{(n,m)}\;=\;A_{++} \Omega^n \omega^m + A_{-+} \Omega^{-n}
\omega^m + A_{+-} \Omega^{n}\omega^{-m} + A_{--}
\Omega^{-n}\omega^{-m}\;.
\end{equation}
Parameters $A_{\pm,\pm}$, related to the eigenvectors of
$\mathcal{M}$, are defined by
\begin{equation}\label{A2}
\Omega^{N-1} A_{+-} \;=\; \omega^{M-1} A_{-+}\;, \quad A_{--}\;=\;
\Omega^{N-1}\omega^{M-1} A_{++}
\end{equation}
and
\begin{equation}\label{A1}
\frac{A_{+-}+A_{--}}{A_{++}+A_{-+}}\;=\;\frac{1-q^2\omega^2}{q^2-\omega^2}\;,\quad
\frac{A_{-+}+A_{--}}{A_{++}+A_{+-}}\;=\;\frac{1-q^2\Omega^2}{q^2-\Omega^2}\;.
\end{equation}
The state (\ref{2wave}) has the evident $N\leftrightarrow M$
symmetry:
\begin{equation}
\boldsymbol{\tau}_\alpha |\Psi\rangle\;=\;
(\omega+\omega^{-1})|\Psi\rangle\;,\quad
\boldsymbol{\tau}_\beta|\Psi\rangle \;=\;
(\Omega+\Omega^{-1})|\Psi\rangle\;.
\end{equation}
From the point of view of $\boldsymbol{\tau}_\beta$, the second
relation of (\ref{BAE2}) is the characteristic polynomial, while
the first one comes from the $\beta$-boundary conditions.

\subsection{$p$-bosons state}

Turn finally to the state with $\mathcal{N}=p$ bosons. Actually,
all the results of this section were obtained explicitly for
$p=3,4$, and then conjectured for arbitrary $p$.

The wave function $\Psi_{j_1,j_2,\dots, j_p}$ of the eigenstate
\begin{equation}
|\Psi\rangle \;=\; \sum_{j_1\leq j_2 \leq \dots \leq j_p}
\Psi_{j_1,j_2,\dots,j_p} \yop_{j_1}\yop_{j_2}\cdots \yop_{j_p}
|0\rangle
\end{equation}
is the superposition of the plane waves. In the generic point (no
coincidence  in $j_1,j_2,\dots,j_p$) the wave function is
\begin{equation}\label{3dba}
\Psi_{j_1,j_2,\dots,j_p} \;=\; \sum_{\sigma,\sigma'} \;
A_{\sigma',\sigma} \;
\Omega_{\sigma_1'}^{n_1}\Omega_{\sigma'_2}^{n_2}\cdots
\Omega_{\sigma'_p}^{n_p} \;
\omega_{\sigma_1^{}}^{m_1}\omega_{\sigma^{}_2}^{m_2} \cdots
\omega_{\sigma^{}_p}^{m_p}
\end{equation}
where $\sigma$ and $\sigma'$ are independent permutations of the
set $(1,2,\dots,p)$, and $n_c,m_c$ are related with $j_c$ by
\begin{equation}\label{double2}
j_c\;=\;Mn_c+m_c\;,\quad 0\leq m_c <M\;,\quad 0\leq n_c < N\;,
\end{equation}
cf. (\ref{double}). Two sets of exponential momenta
\begin{equation}\label{momenta}
\omega_c=\EXP^{ik_c^{}}\;,\quad \Omega_c=\EXP^{ik_c'}\;,\quad
c=1,2,\dots, p\;,
\end{equation}
are solutions of two sets of equations. To write out these
equation, we need some extra notations. Firstly, let
\begin{equation}\label{G}
G_{a,b}(\{\omega\}) \;=\;
\frac{q^{-1}\omega_b-q\omega_a}{\omega_b-\omega_a}\;.
\end{equation}
Let $I_k$ be a length-$k$ subsequence of $(1,2,\dots,p)$,
$I_{p-k}$ be the compliment subsequence such that
\begin{equation}
I_k \cup I_{p-k} = (1,2,\dots,p)\;.
\end{equation}
Let then
\begin{equation}\label{P}
P_k[\{\omega,\omega^M\}]\;=\; \sum_{I_k} \left(\prod_{a\in I_k}
\omega_a^{M}\prod_{b\in I_{p-k}} G_{b,a}(\{\omega\})\right)\;,
\end{equation}
where the sum is taken over all possible subsequences of the
length $k$. E.g. for $p=3$
\begin{equation}
\begin{array}{l}
\ds P_1[\{\omega,\omega^M\}] = \omega_1^M G_{21} G_{31}+\omega_2^M
G_{12} G_{32}+\omega_3^M G_{13} G_{23}\;,\\
\ds P_2[\{\omega,\omega^M\}] = \omega_1^M\omega_2^M G_{32}
G_{31}+\omega_1^M\omega_3^M
G_{21} G_{23} + \omega_2^M\omega_3^M G_{12} G_{13}\;,\\
\ds P_3[\{\omega,\omega^M\}]=\omega_1^M\omega_2^M\omega_3^M\;.
\end{array}
\end{equation}
The final step: let
\begin{equation}\label{Poly}
\mathfrak{P}(\Omega|\{\omega,\omega^M\})\;=\;\sum_{k=0}^{p} (-)^k
\Omega^{p-k} P_k[\{\omega,\omega^M\}]\;=\; \Omega^{p} -
\Omega^{p-1}P_1 + \Omega^{p-2}P_2+\cdots\;.
\end{equation}
Then the consistency conditions for the Ansatz (\ref{3dba}) read
\begin{equation}\label{3dbae}
\left\{\begin{array}{l} \ds
\mathfrak{P}(\Omega_a|\{\omega,\omega^M\})\;=\;0\;,\\
\\
\ds \mathfrak{P}(\omega_a|\{\Omega,\Omega^N\}) \;=\; 0\;,
\end{array}\right. \quad \forall\quad a=1,2,\dots,p\;.
\end{equation}
These are the Bethe Ansatz equations for our two dimensional Bose
gas.

The values of $\Psi_{j_1,j_2,\dots,j_p}$ in the case when some of
$j_a$ coincide, as well as the values of $(p!)^2$ amplitudes
$A_{\sigma,\sigma'}$, can be defined uniquely.

The state (\ref{3dba}) provides the eigenvalues of (\ref{norm})
\begin{equation}
\boldsymbol{\tau}_\alpha\; |\Psi\rangle \;=\; |\Psi\rangle \;
\left(\sum_{a=1}^p \omega_a^{-1}\right)\;,\quad
\boldsymbol{\tau}_\beta\; |\Psi\rangle \;=\; |\Psi\rangle \;
\left(\sum_{a=1}^p \Omega_a^{-1}\right)\;.
\end{equation}
Note,
\begin{equation}\label{mass-center}
\Omega_1\Omega_2\cdots \Omega_p\;=\;\omega_1^M\omega_2^M\cdots
\omega_p^M
\end{equation}
(it follows from $P_p=\omega_1^M\cdots \omega_p^M$), therefore the
mass center of $p$-particles state moves accordingly to
(\ref{k-1},\ref{tr-1}).

The one-dimensional limit of (\ref{3dbae}) must be mentioned. In
the case $N=1$ our system becomes the chain of the length $M-1$.
Polynomial $\mathfrak{P}(\omega|\{\Omega,\Omega\})$ has a simple
structure, all $\Omega_a$ may be excluded from (\ref{3dbae}). The
resulting equations for $\omega_a$ are
\begin{equation}
\omega_a^{M-1}\;=\;\prod_{b\neq a}\;
\frac{G_{ab}(\{\omega\})}{G_{ba}(\{\omega\})}\;,
\end{equation}
what are the usual Bethe Ansatz equations for the quantum chain.

The last important note is that equations (\ref{3dbae}) provide
real momenta $k_c^{}$ and $k_c'$ of (\ref{momenta}) if $q$ is
real. In addition, numerical estimations for not too big lattices
and for two particles shows that the Ansatz is complete.

\subsection{Sketch derivation of (\ref{3dbae})}

Equations (\ref{3dbae}) may be obtained in the same way as
(\ref{BAE2}). One may start with the case when $m_1< m_2 < \dots <
m_p$ and $m_p<m_1+M$. Then the eigenstate of
$\boldsymbol{\tau}_\alpha$, $\ds \boldsymbol{\tau}_\alpha \;
|\Psi\rangle = |\Psi\rangle\tau_\alpha$,
$\ds\tau_\alpha=\sum_a\omega_a^{-1}$, is given by
\begin{equation}
\Psi_{m_1,...,m_p} \;=\; \sum_{\sigma} A^{(0)}_\sigma
\omega_{\sigma_1}^{m_1}\omega_{\sigma_2}^{m_2}\cdots
\omega_{\sigma_p}^{m_p}\;.
\end{equation}
The eigenvalue equation for $\boldsymbol{\tau}_\alpha$ provide the
way to interpolate $\Psi_{m_1,m_2,\dots, m_p}$ for the larger
values of $m_a$. For instance, when $m_p\sim m_1+M$, the
eigenvalue equation modifies as
\begin{equation}
\tau_\alpha\Psi_{m_1,\dots,m_p}=q^{\delta_{m_p,m_1+M-1}}\Psi_{m_1-1,\dots,m_p}
+q^{\delta_{m_p,m_1+M}}\Psi_{m_1,\dots,m_p-1} + \textrm{all the
rest}\;.
\end{equation}
Extra $q$-factors produce some linear transformation of the
amplitudes $A_\sigma$ of $\Psi_{m_1,\dots,m_p}|_{m_p\geq m_1+M}$
in the some way as in eq. (\ref{PQ-matrix}). Repeating this
procedure for $m_p\sim m_2+M$, $m_p\sim m_3+M$ etc., one comes to
\begin{equation}\label{plusM}
\Psi_{m_1,m_2,\dots, m_p+M}\;=\;\sum_\sigma A_{\sigma}^{(1)}
\omega_{\sigma_1}^{m_1}\omega_{\sigma_2}^{m_2} \cdots
\omega_{\sigma_p}^{m_p}\;,
\end{equation}
where $A_{\sigma}^{(1)}$ is a linear combination of
$A_{\sigma}^{(0)}$:
\begin{equation}
A_{\sigma}^{(1)} \;=\; \sum_{\sigma'} \mathcal{M}_{\sigma,\sigma'}
\; A_{\sigma'}^{(0)}\;,
\end{equation}
cf. (\ref{PQ-matrix}). The Ansatz (\ref{3dba}) corresponds to
(\ref{plusM}) in the basis of eigenvectors of $\mathcal{M}$. The
miracle of the exact integrability is that $p!\times p!$ matrix
$\mathcal{M}$ has only $p$ eigenvalues:
\begin{equation}\label{miracle}
\det \left( \Omega - \mathcal{M}\right) \;=\;
\mathfrak{P}(\Omega|\{\omega,\omega^M\})^{(p-1)!}
\end{equation}
The second equation in (\ref{3dbae}), providing the
$N\leftrightarrow M$ symmetry of the Bethe Ansatz equations,
solves the boundary condition analogously to (\ref{MPQ}).

\section{Conclusion}

This paper presents a model of two-dimensional Bose gas and the
conjecture for the equations (\ref{3dbae}) describing its
eigenstates. The model has the main features of a physical model.
The states are described by the system of plane waves -- at least
the notion of the momenta is well defined. In particular, the sum
of cosines of momenta is a candidate for the Hamiltonian.
Equations (\ref{3dbae}), in the case when the number of bosons $p$
is big, are the subject of further investigations. We believe,
this model may have some interest for the theory of integrable
systems and for the condensed matter physics.

\noindent\textbf{Acknowledgements.} I would like to thank M.
Batchelor, V. Bazhanov, M. Bortz, X.-W. Guan and V. Mangazeev for
valuable discussions.


\begin{thebibliography}{1}

\bibitem{Sergeev:2000tmp}
S.~M. Sergeev.
\newblock {A}uxiliary transfer matrices for three-dimensional integrable
  models.
\newblock {\em Theor. Math. Phys.}, 124:391--409, 2000.

\bibitem{Sergeev:2005b}
V.~V. Bazhanov and S.~M. Sergeev.
\newblock {Z}amolodchikov's {T}etrahedron {E}quation and {H}idden {S}tructure
  of {Q}uantum {G}roups.
\newblock {\em hep-th/0509181}, 2005.

\bibitem{Sergeev:1999jpa}
S.~M. Sergeev.
\newblock {Q}uantum {$2+1$} evolution model.
\newblock {\em J. Phys. A: Math. Gen.}, 32:5693--5714, 1999.

\bibitem{Sergeev:2005d}
S.~M. Sergeev.
\newblock {Q}uantum curve in q-oscillator model.
\newblock {\em nlin.SI/0510048}, 2005.

\bibitem{Sergeev:2004pn}
S.~M. Sergeev.
\newblock {Q}uantum integrable models in discrete 2+1 dimensional space-time:
  auxiliary linear problem on a lattice, zero curvature representation,
  isospectral deformation of the {Z}amolodchikov-{B}azhanov-{B}axter model.
\newblock {\em Particles and Nuclei}, 35:1051--1111, 2004.

\bibitem{Bazhanov:1990qk}
V.~V. Bazhanov, R.~M. Kashaev, V.~V. Mangazeev, and Yu.~G.
Stroganov.
\newblock ${Z}_n^{\otimes(n-1)}$ generalization of the chiral {P}otts model.
\newblock {\em Commun. Math. Phys.}, 138:393--408, 1991.

\bibitem{Date:1990bs}
Etsuro Date, Michio Jimbo, Kei Miki, and Tetsuji Miwa.
\newblock Generalized chiral {P}otts models and minimal cyclic representations
  of ${U}_q(gl(n,{C}))$.
\newblock {\em Commun. Math. Phys.}, 137:133--148, 1991.

\bibitem{Bazhanov:1981zm}
V.~V. Bazhanov and Yu.~G. Stroganov.
\newblock On commutativity conditions for transfer matrices on multidimensional
  lattice.
\newblock {\em Theor. Math. Phys.}, 52:685--691, 1982.

\bibitem{Sergeev:2005c}
S.~M. Sergeev.
\newblock {I}ntegrability of {$q$}-oscillator lattice model.
\newblock {\em nlin.SI/0509043}, 2005.

\end{thebibliography}

\end{document}